\documentclass[aps,prl,reprint,superscriptaddress,amsmath]{revtex4-1}
\newcommand{\be}{\begin{equation}}%\extraspace}
\newcommand{\ee}{\end{equation}}
\usepackage{graphicx}
\usepackage{dcolumn}
\usepackage{bm}
\usepackage{color}
\begin{document}

%\preprint{}

\title{Single parameter scaling of one-dimensional systems with real-space long-range correlated disorder }


\author{Greg M. Petersen}
\author{Nancy Sandler}
%\thanks{}
\affiliation{Department of Physics and Astronomy, Nanoscale
   and Quantum Phenomena Institute, and Condensed Matter and Surface Science Program,\\Ohio University, Athens, Ohio
   45701-2979}
\date{\today}

\begin{abstract}
Advances in material growth methods have renewed the interest in localization of one-dimensional systems in the presence of scale-free long-range correlated disorder potentials. We analyze the validity of single parameter scaling for the $\beta$ function away from the band center, in the presence of correlations. A renormalized disorder strength emerges reducing the regime of validity of the single parameter scaling hypothesis.  Analysis of localization lengths and participation ratios leads to correlation dependent critical and fractal exponents, consistent with the extended Harris criterion.
\end{abstract}

\pacs{73.20.Fz, 73.20.Jc, 73.63.Nm,73.22.Dj}

\maketitle

The Anderson model of disordered one-dimensional (1d) systems provides valuable insight into the electronic transport properties of a wide variety of materials such as DNA\cite{Lewis, Klotsa, Qu}, polymers\cite{Hjort, PrigodinI, PrigodinII}, random media\cite{Wang} and nanowires\cite{Pascual}, for which the scaling theory of localization assumes exponentially localized eigenstates with vanishing conductance\cite{Abrahams}. Historically, the successful application of scaling theory to systems with short-range correlations\cite{50years} was followed by the development of correlated disorder models\cite{KramerB:1986}. These models fall into three basic categories: those with quasi-periodic real space order\cite{Jitomirskaya,Avila}, random disorder amplitudes chosen from a discrete set of values\cite{Flores,Phillips}, and models with specific long-range correlations\cite{DeMoura}.  In the first two categories, a discrete number of eigenstates are predicted to remain extended. The last class of models however, produce a mobility edge that separates regions of extended and localized states in the Hilbert space, rendering a metal-insulator transition in one-dimension.  As a consequence, much work has been devoted in the last few years to understand the effects of different long-range correlations and, in particular, those with power-law spectral functions, generated by self-affine potentials\cite{Izrailev}.  Disorder potentials of this class however are not easily realizable in controlled experimental setups, thus several authors have focused on potentials with power-law correlations in real-space. These have already appeared in experiments in ultra-cold atom systems\cite{Billy,Inguscio}, where localization physics was achieved using speckle potentials, known to possess long-range correlations. Furthermore, advances in materials growth methods such as molecular self-assembly on surfaces or wires\cite{Saw1, Saw2}, and the production of rippled graphene flakes and ribbons\cite{Novoselov,Lau,Wang-Y}, makes it possible to obtain systems where real-space correlated disorder potentials can be manipulated, making the understanding of these models of experimental relevance.

Transfer matrix studies for real-space power-law correlated models have shown, for all disorder strengths: i) complete localization for all values of the power-law exponent $\alpha$, ii) anomalous enhancement of localization for band edge states, iii) more extended band center states as compared to those in the presence of short-range correlated potentials\cite{Russ, CroyA}.  Furthermore, a scaling analysis of a weak disorder expansion predicts a rescaling of the disorder strength $W \rightarrow W_{eff}$ that takes place near the band center, where $W_{eff}=\sqrt{S(2k)}W$\cite{Krokhin}, and $S(2k)$ is the $2k$ component of the spectral density associated to the correlated potential. Finally, it was shown that a white-noise disorder model reproduces the effect on the band edges\cite{CroyA}, suggesting a different scaling behavior of the potential amplitudes in this energy region. Note that these studies have not addressed the anomalies at the band edge and centers found in studies with uncorrelated potentials\cite{Schomerus,Deych,Kravtsov}

Although it is well established by now that all eigenstates are localized for these correlations, other effects with direct relevance for applications have not been fully addressed up to date. In this work, we focus on three of them: i) validity of single parameter scaling (SPS), ii) crossover between the two predicted scaling regimes, and iii) nature of localized states as described by the localization length critical exponent $\nu$ and the participation ratio fractal exponent $D$. 

First, we analyze the single parameter scaling hypothesis for the $\beta$ function (an issue that remains controversial at present\cite{Izrailev}).  Second, we calculate the localization length $\xi$ and study the crossover that separates eigenstates into two regions: those affected by $W_{eff}$ and those affected by an effective white-noise disorder.  We characterize it by analyzing the change in $\nu$. Finally we calculate the inverse participation ratio and obtain the corresponding values for $D$ for localized wave functions. Our results suggest that $\nu$ and $D$ acquire a dependence on the correlated disorder power-law exponent $\alpha$, consistent with predictions by the extended Harris criterion\cite{Weinrib,Sandler}.

\emph{Model}:
We begin with the Anderson hamiltonian for a one-dimensional chain of $N = L/a$ sites in real space, with $L$ the length of the chain and $a$ the lattice constant:
\begin{equation}
\label{eq:tb}
H = \sum_{n=0}^{N-1} \epsilon_n c^{\dag}_n c_n -  t( c^{\dag}_n c_{n+1} + h.c.)
\end{equation}
Here $\epsilon_n$ is  the random on-site energy , $t$ is the hopping energy, and $c^{\dag}_n$ ($c_n$) is the fermion creation (annihilation) operator at site $n$.  $\epsilon_n$ is obtained from a gaussian distribution satisfying the conditions $\langle \epsilon_n \rangle = 0$,  $\langle \epsilon^2_n \rangle = W$, and with a normalized correlation function
\begin{equation}
\label{acf}
\Gamma_n = \frac{\langle \epsilon_n \epsilon_0 \rangle}{\langle \epsilon_0^2 \rangle} = \frac{1}{(1+n)^\alpha}
\end{equation}
where $n$ is the distance between sites, and $\alpha$ determines the correlation strength.  Although several choices render the same asymptotic form $1/n^\alpha$ as $n \to \infty$\cite{CroyA}, short range details are irrelevant in the thermodynamic limit which are the focus of this work.  

\emph{Methods}:  To generate a stochastic set of real on-site energies, a set of random complex numbers in reciprocal space (k-space) is generated, with the condition $\tilde{\epsilon}_k = \tilde{\epsilon}_{-k}^*$ ($\tilde{\epsilon_k}$ is the complex Fourier component of the real space energy $\epsilon_{n}$).  These variables are chosen from a gaussian distribution with a standard deviation equal to the spectral density $S(k) = \langle |\tilde{\epsilon}_k|^2 \rangle$.  A Fourier transform of Eq.($\ref{acf}$), renders $S(k)$ in a computationally friendly form:
\be
\langle |\tilde{\epsilon}_k|^2 \rangle = \frac{1}{4N^2}\Big(2N + Re\Big[F\Big(\frac{2n}{(1+2N-n)^\alpha}\Big)\Big]\Big)
\ee
Here, $F()$ and $Re()$ denote the Fourier transform and real part of the argument respectively. The discrete spectral density is obtained with a Fast Fourier Transform (FFT) algorithm, a random set \{$\tilde{\epsilon_k}$\} is generated, and the set of \{$\epsilon_{n}$\} is obtained via the inverse Fourier transform.  

\begin{figure}
\begin{center}
\includegraphics[scale=.48]{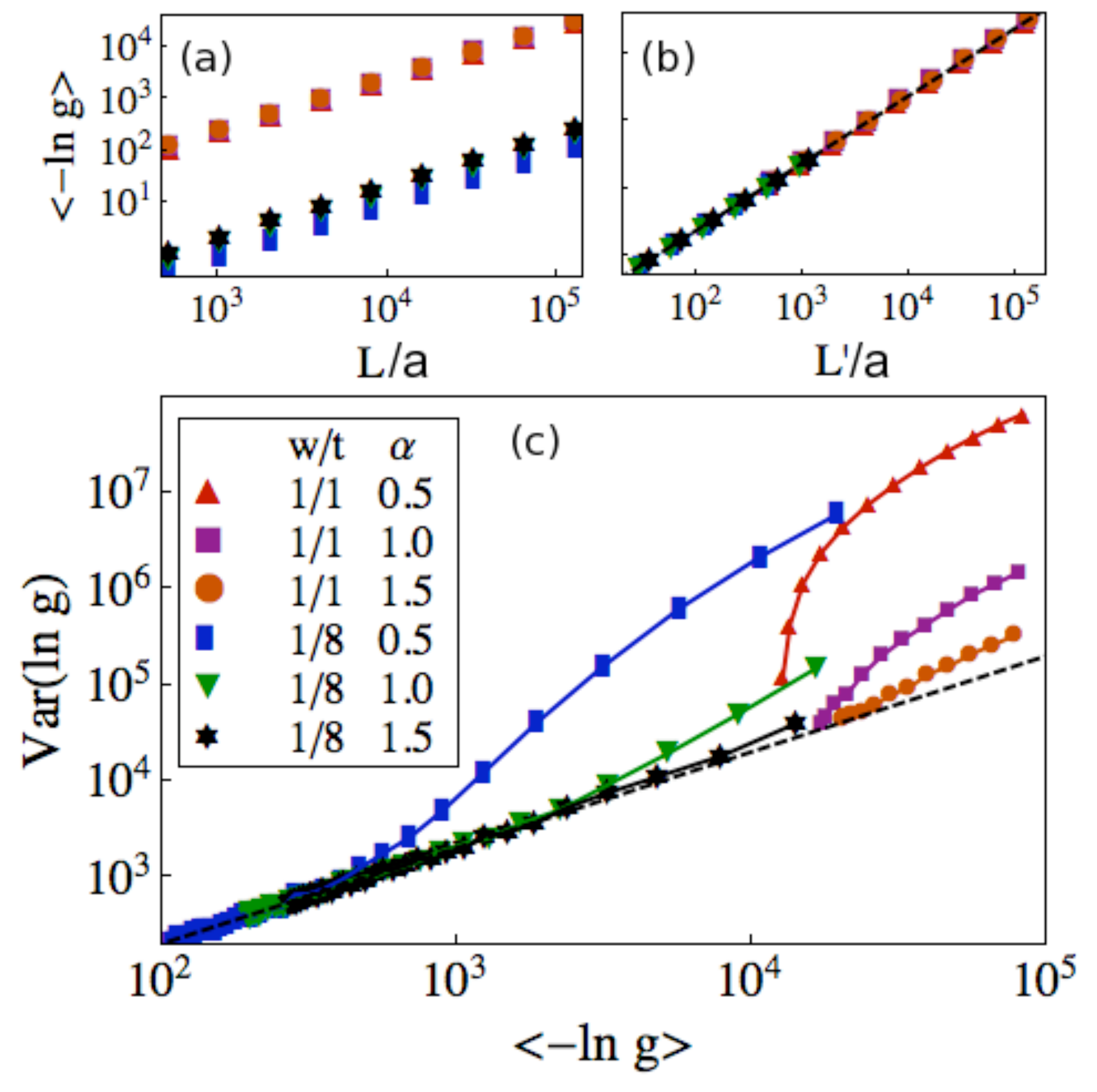}
\caption{Scaling of the average value of the typical conductance and it's relation to the variance. Data is averaged over 1000 disorder configurations.  Symbol size represents statistical error. (a) Raw conductance data for $L/a = 2^{9}$-$2^{17}$ and various parameters. (b) Data collapse by rescaling of $L$ by $\xi$ as described in the text.  Notice the universal slope for all values in a.) (c) Relationship between variance and average for L/a=$2^{17}$.  Dashed black line shows Var$(\ln g) = 2 \langle -\ln g \rangle$.}
\label{Lng}
\end{center}
\end{figure}

\emph{Scaling function}: The validity of the SPS hypothesis for the dimensionless conductance of systems with long-range correlated potentials is still an open question\cite{Izrailev}.  For example, for the disorder potential considered in Ref.\cite{DeMoura}, the exponent determines an effective length scale beyond which the real-space correlation function becomes negative\cite{Greg2,Greg3}. To answer the question, for $1/n^\alpha$ correlated potentials, we use the relation Var$(\ln g) = -2 \langle \ln g \rangle$ between the first (average) and second (variance) cumulants of the "typical" conductance $\ln g$, found to hold for uncorrelated disorder potentials\cite{Shapiro,Beenakker}, as the criterion for the validity of SPS.

The dimensionless conductance $g$ is equal to the transmission function $T$ obtained from advanced and retarded Green's functions by $T = Tr(\Gamma_L G^r_c \Gamma_R G^a_c)$ where:
\be
G^{r,a}_c (E)= \frac{1}{E-H _d- \Sigma^{r,a}_L - \Sigma^{r,a}_R}.
\ee
Here $E$ is the energy, $H_d$ the hamiltonian of the disordered system, and $\Sigma^{r,a}_{L,R}$ self energies due to left and right leads.  The hybridization functions $\Gamma_L$ and $\Gamma_R$  are defined by $\Gamma_{L,R} = i[\Sigma^r_{L,R} - \Sigma^a_{L,R}]$.

Usually $\Sigma^{r,a}_{L,R}$ is found numerically through decimation procedures\cite{LakeR}, however for semi-infinite one-dimensional leads, an analytic expression can be derived\cite{Arrachea}:
\be
\Sigma^{r,a} = \frac{2t^2}{E \pm i\sqrt{4t^2 - E^2}}.
\ee
To calculate the inverse of $G^{r,a}_{c}(E)$  we use recursive methods that require an order $\sim N$ of operations \cite{RGF}, allowing to maximize the number of disorder realizations and system sizes.  Results from these calculations are plotted in Fig.~\ref{Lng}.  Panel (a) and (b) show the collapse of each curve after a rescaling of $L$ by $\xi$ (with $\xi = \lim_{L \to \infty}\frac{-\ln g}{L}$), indicating $\frac{d \langle \ln g \rangle}{d \ln L} = \langle \ln g \rangle$.  The value of the $\beta$ function is negative confirming localization physics. Panel (c) shows the relation between the first two cumulants. The data suggests SPS is valid for low energies but violated near the band edge for $\frac{W}{t\alpha} < 1$ and for all energies if $\frac{W}{t\alpha} \ge 1$.

\begin{figure}
\begin{center}
\includegraphics[scale=0.45]{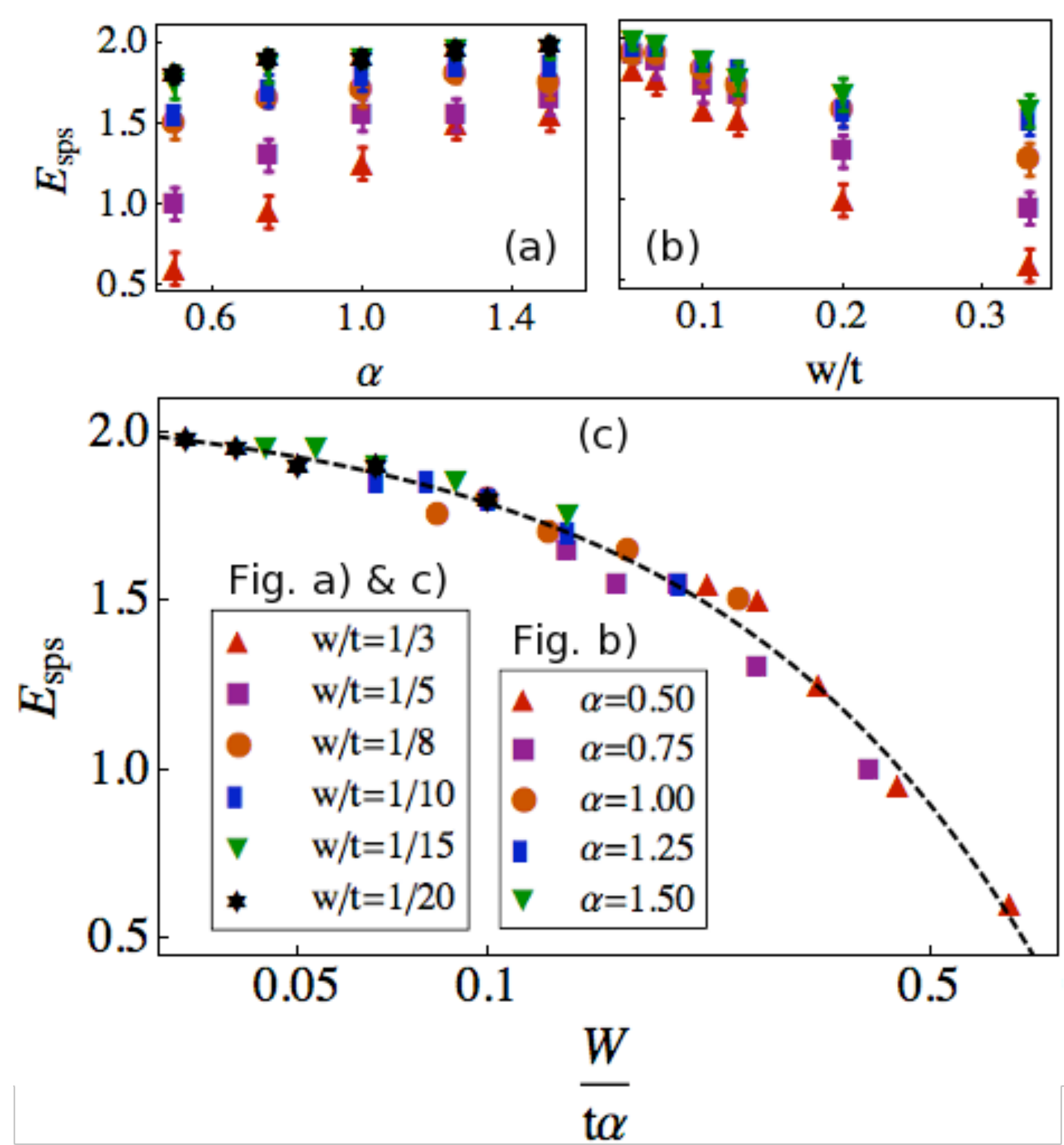}
\caption{Energy $E_{sps}$ separating valid ($E < E_{sps}$) and violated ($E >  E_{sps}$) SPS regions as function of $\alpha$ (panel (a)) and disorder strength (panel (b)). Panel (c) shows $E_{sps}$ as function of the rescaled variable $\frac{W}{t\alpha}$.  The black dashed line represents the fit of Eq. \ref{Espsfit}.}
\label{Varg}
\end{center}
\end{figure}

We define $E_{sps}$ as the energy where SPS is first violated, for each value of $\alpha$ and $W/t$. Fig.~\ref{Varg} shows the dependence of  $E_{sps}$ on $\alpha$ (a) and $W/t$ (b).  By defining a rescaled variable $\frac{W}{t\alpha}$, all curves collapse as shown in (c).  Numerical fits of the data lead to the expression
\begin{equation}
E_{sps} = 2.1 -(\frac{5}{2}\frac{W}{t\alpha})^{0.84}.
\label{Espsfit}
\end{equation}
reminiscent of the relation found in Ref.\cite{Derrida} for the energy at which the perturbation expansion of $\xi(E)$ done with uncorrelated potentials near $E\sim 2$ ceases to be valid\cite{Derrida}. 

\begin{figure}
\begin{center}
\includegraphics[scale=0.45]{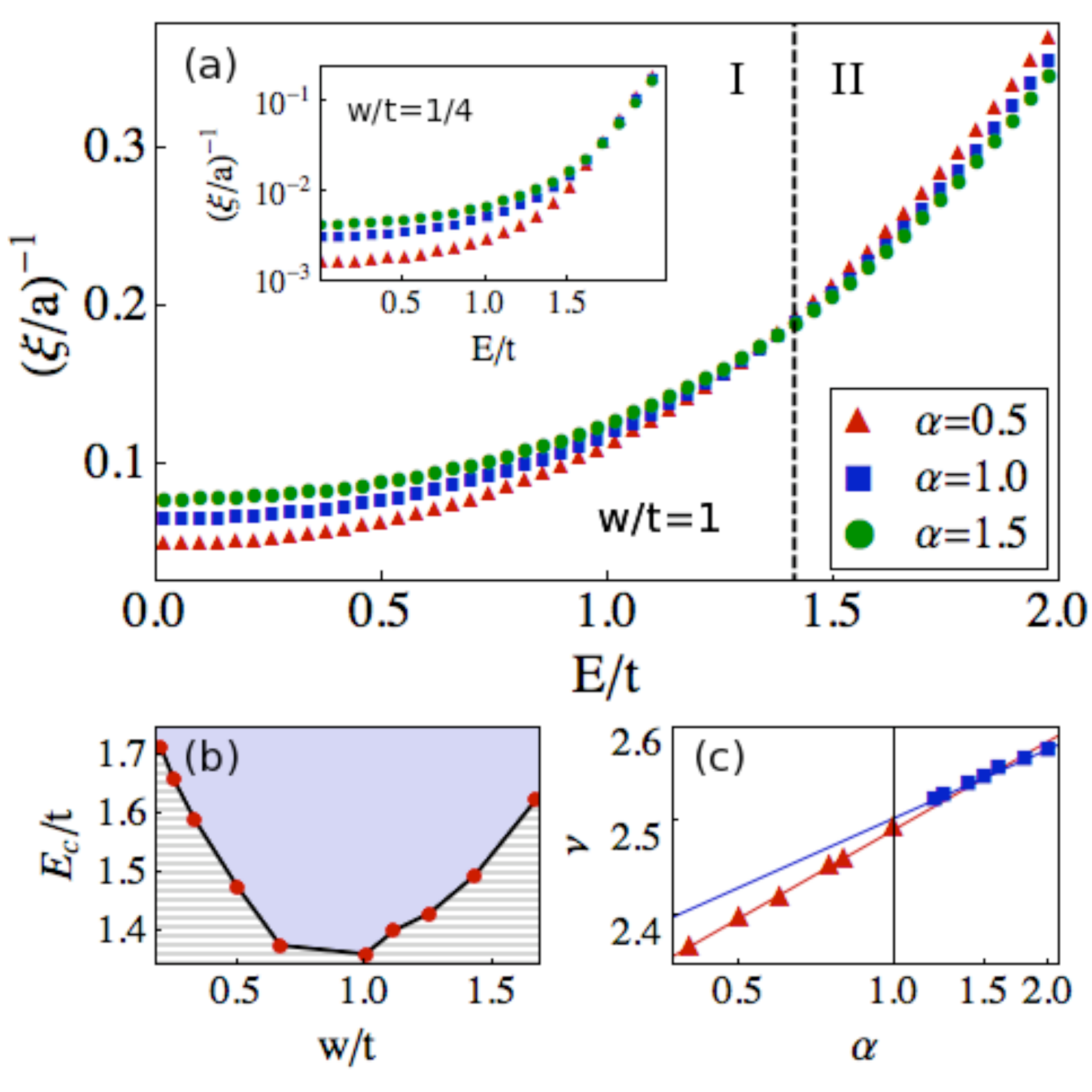}
\caption{(a) Localization length vs energy for system sizes $L/a = 2^{11}$-$2^{19}$ and 1000 disorder configurations.  Eigenstates become more extended with decreasing values of $\alpha$ in region I. At $E_{c}/t \sim \sqrt{2}$ (dashed line), $\xi$ is independent of $\alpha$.  In region II, there is enhanced localization.  (Positive values of E are plotted only). (b) Dependence of crossing energy $E_{c}/t$ vs disorder strength $W/t$.  Solid, colored region corresponds to energies with enhanced localization while dashed region corresponds to more extended-like states.  (c) Localization length critical exponent $\nu$ as a function of $\alpha$ for $W/t =1$ showing different functional dependence for $\alpha<$ and $>1$. Symbol sizes represent statistical error bars.}
\label{LocVEt1}
\end{center}
\end{figure}

\emph{Localization Length}:  Next, we apply transfer matrix techniques  to obtain $\xi$  as a function of energy \cite{Russ, CroyA} by calculating the lowest Lyapunov exponent from the successive multiplication of a position dependent transfer matrix
\be
T_i = \left( \begin{matrix} (E-\epsilon_i)/t&-1\\ 1&0 \end{matrix}\right).
\ee
We show a typical data set in Fig.~\ref{LocVEt1}(a). The values  plotted are obtained by calculating $\xi$ for different systems sizes and extrapolating to $L \to \infty$. As predicted from scaling arguments, and partially confirmed in transfer matrix calculations\cite{CroyA, Krokhin}, two regions are distinguished with enhanced and suppressed localization. These regions are separated at the crossing energy $E_c/t \sim \sqrt{2}$, defined as the energy where $\xi$ is independent of $\alpha$ for a fixed value of $W/t$. These results indicate that correlations produce more extended states in general (i.e. the region of 'band-center' states extends closer to the band-edge) with localized states 'squeezed' into the band edges. They also suggest a continuos enhancement of extended states as $\alpha$ is reduced with the singular limit of perfect extended states for $\alpha = 0 $. The energy $E_{c}$, that depends on the disorder strength as shown in Fig.~\ref{LocVEt1}(b), separates the region of validity for the two scaling regimes proposed in Refs.~\cite{Krokhin, Russ}.  A quantitative analysis of the critical exponent $\nu$ obtained from $\xi \sim |E-E_{c}|^{-\nu}$ (with $E_{c}=0$ in this model), produces the results shown in  Fig.~\ref{LocVEt1}(c) for $W/t =1$. The data suggests $\nu=\nu(\alpha)$ with a functional dependence that changes at $~\alpha =1$, consistent with the extended Harris criterion for classical \cite{Weinrib} and quantum systems \cite{Sandler}. Similar behavior is obtained for other values of $W/t$ not shown.

\emph{Participation ratio}: To get insight into the nature of localized eigenstates we performed an analysis of the participation ratio. This quantity\cite{Thouless} provides information about real space fluctuations of eigenstate amplitudes. For a state $m$ it is defined by:
\begin{equation}
A_m = \frac{1}{N}\frac{\Big(\sum_{n=0}^{N-1} |\psi_m(n)|^2 \Big)^2}{\sum_{n=0}^{N-1} |\psi_m(n)|^4}
\end{equation}
where $\psi_m(n)$ denotes the amplitude of the eigenstate at position $n$.  If the state is localized, $A_{m}$ approaches zero as $A_m \sim \frac{1}{(L/a)^D}$ as the system size increases. Here $D$ is the fractal exponent, that  characterizes the fragmented nature of localized eigenstates\cite{Roman1,Roman2,Evers}. 

\begin{figure}
\begin{center}
\includegraphics[scale=.54]{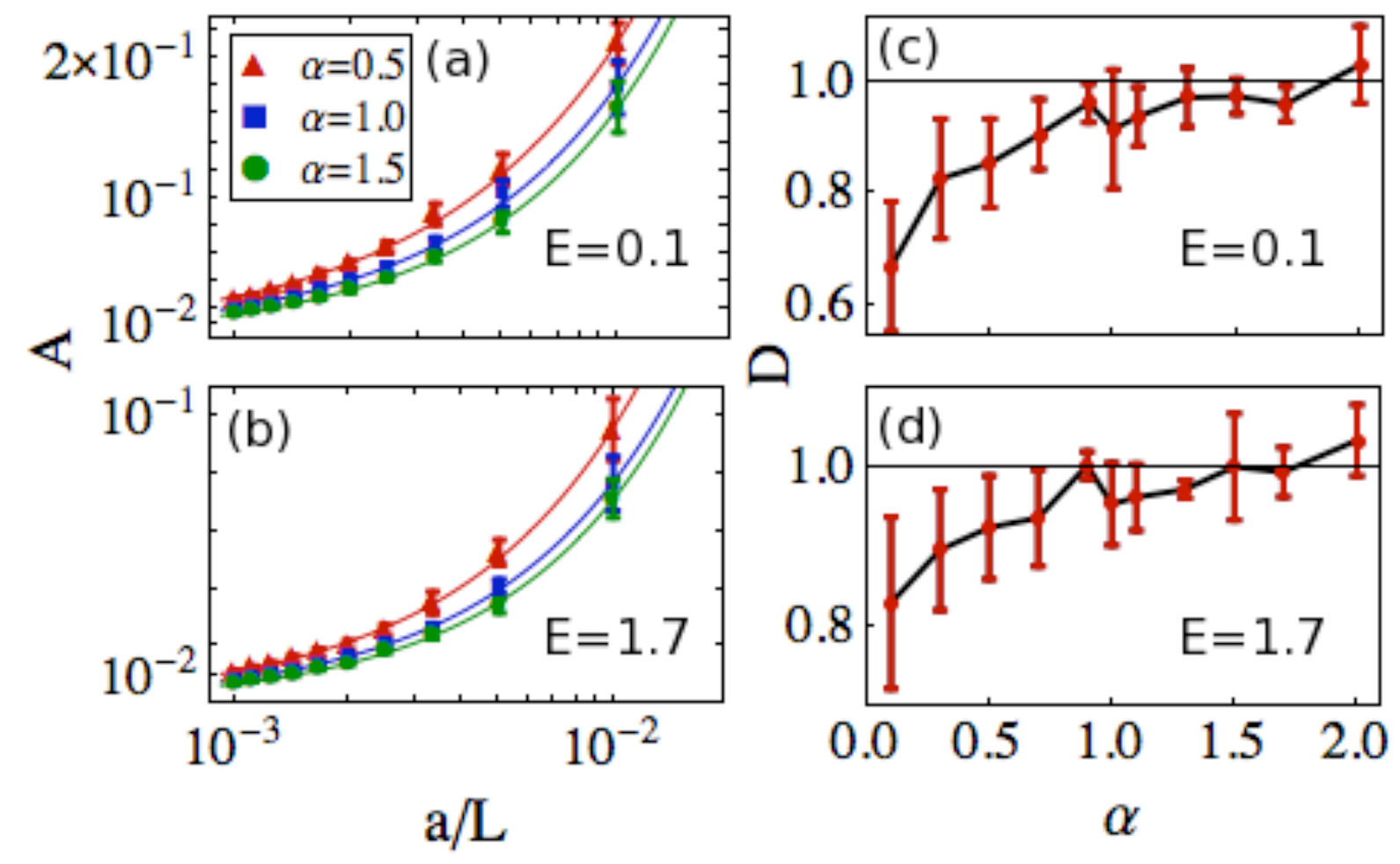}
\caption{Participation ratio $A_{m}$ for states at energies $E/t$ = 0.1 (a) and 1.7 (b) for $L/a$ = 100-1000 averaged over 100 disorder realizations (error bars obtained from disorder averaging) and $W/t$=1. Largest value for $\xi/a \sim 30$.  Solid lines are fitting functions of the form $A = C(a/L)^D$. Error bars are obtained by error propagation from proposed fitting function.  Panels (c) and (d) show values of fractal exponent $D$  for different values of $\alpha$.  A change at $\alpha=1$ is observed, consistent with the SPS results.}
\label{PRFit}
\end{center}
\end{figure}

An exact diagonalization of Eq.~(\ref{eq:tb}) renders the eigenstates and the numerical evaluation of $A_{m}$ as a function of system size.  Fig.~\ref{PRFit} shows results for two different states, chosen near the band center (a) and the band edge (b) respectively for $W/t=1$ as representative data.  Continuous lines correspond to a fitting function $c_o(L/a)^{-D}$ with $D$ and $c_o$ fitting parameters dependent on $W/t, E/t$ and $\alpha$. In panels (c) and (d), the fractal dimension $D$ is plotted versus the correlation length exponent $\alpha$.  The data suggests the existence of two regimes for $\alpha >1$ and $\alpha < 1$. In the first regime $D$ is independent of $\alpha$, a result expected for uncorrelated disorder potentials \cite{Roman2}. However, for $\alpha <1$, $D$ acquires an $\alpha$-dependence that appears stronger for band-center states than for band-edge states. The decrease of $D$ with $\alpha$ indicates more extended eigenstates (less fragmented regions with non-zero wave function amplitude) near the band center. The particular dependence of $D$ with $\alpha$ is also reminiscent of the relation found between $\nu$ with $\alpha$ in the previous section.

\emph{Conclusions:} We have shown that in the presence of real space scale-free correlated disorder potentials with power-law (positive) correlations, the SPS hypothesis is violated for energies between the band edge and center, when the ratio $\frac{W}{t\alpha} < 1$.  For $\frac{W}{t\alpha} > 1$, single parameter scaling is violated for all energies.  The $\alpha$-dependent energy threshold at which SPS is violated, $E_{sps}$, is consistent with the value for which the effects of the band-edge anomaly appear in the expansion for $\xi$ in terms of $W/t$, thus linking the SPS violation to the non-analyticity of $\xi$. While this could suggest that correlated potentials are equivalent to uncorrelated ones with a rescaled disorder strength $W' \sim W/\alpha$, the anomalous increase of the localization length for states with energies $E \leq E_{c}$ is inconsistent with this picture (with $E_{c}$ the crossover energy between two scaling regimes). In fact, correlations affect the nature of localized states depending on their energy, as shown by the variation of $\nu$ and $D$ as functions of $\alpha$. In particular, they produce an increased number of extended-like states for $E \leq E_{c}$, more pronounced when $\alpha \leq 1$, indicating a continuous interpolation to perfectly extended states in the limit $\alpha = 0 $. The clear departure from SPS with the appropriate degree of correlations, ubiquitous in a variety of real systems, calls for the need of careful interpretation of experimental data in systems with these types of disorder realizations. 

{\it Acknowledgements}. We appreciate useful discussions with  B. Shapiro, M. Lyra, , S. Russ, E. Mucciolo, A. M. Llois, A. Barral, C. Lewenkopf and C. Mudry. This work was supported by NSF PIRE and MWN/CIAM grants. We acknowledge support from KAVLI Sta. Barbara and Dahlem Center for Complex Systems, FU Berlin, where parts of this work was completed.

\bibliography{MyBib}

\begin{thebibliography}{44}%
\makeatletter
\providecommand \@ifxundefined [1]{%
 \@ifx{#1\undefined}
}%
\providecommand \@ifnum [1]{%
 \ifnum #1\expandafter \@firstoftwo
 \else \expandafter \@secondoftwo
 \fi
}%
\providecommand \@ifx [1]{%
 \ifx #1\expandafter \@firstoftwo
 \else \expandafter \@secondoftwo
 \fi
}%
\providecommand \natexlab [1]{#1}%
\providecommand \enquote  [1]{``#1''}%
\providecommand \bibnamefont  [1]{#1}%
\providecommand \bibfnamefont [1]{#1}%
\providecommand \citenamefont [1]{#1}%
\providecommand \href@noop [0]{\@secondoftwo}%
\providecommand \href [0]{\begingroup \@sanitize@url \@href}%
\providecommand \@href[1]{\@@startlink{#1}\@@href}%
\providecommand \@@href[1]{\endgroup#1\@@endlink}%
\providecommand \@sanitize@url [0]{\catcode `\\12\catcode `\$12\catcode
  `\&12\catcode `\#12\catcode `\^12\catcode `\_12\catcode `\%12\relax}%
\providecommand \@@startlink[1]{}%
\providecommand \@@endlink[0]{}%
\providecommand \url  [0]{\begingroup\@sanitize@url \@url }%
\providecommand \@url [1]{\endgroup\@href {#1}{\urlprefix }}%
\providecommand \urlprefix  [0]{URL }%
\providecommand \Eprint [0]{\href }%
\providecommand \doibase [0]{http://dx.doi.org/}%
\providecommand \selectlanguage [0]{\@gobble}%
\providecommand \bibinfo  [0]{\@secondoftwo}%
\providecommand \bibfield  [0]{\@secondoftwo}%
\providecommand \translation [1]{[#1]}%
\providecommand \BibitemOpen [0]{}%
\providecommand \bibitemStop [0]{}%
\providecommand \bibitemNoStop [0]{.\EOS\space}%
\providecommand \EOS [0]{\spacefactor3000\relax}%
\providecommand \BibitemShut  [1]{\csname bibitem#1\endcsname}%
\let\auto@bib@innerbib\@empty
%</preamble>
\bibitem [{\citenamefont {Lewis}\ \emph {et~al.}(2003)\citenamefont {Lewis}
  \emph {et~al.}}]{Lewis}%
  \BibitemOpen
  \bibfield  {author} {\bibinfo {author} {\bibfnamefont {J.~P.}\ \bibnamefont
  {Lewis}} \emph {et~al.},\ }\href@noop {} {\bibfield  {journal} {\bibinfo
  {journal} {J. of Phys. Chem. B}\ }\textbf {\bibinfo {volume} {107}},\
  \bibinfo {pages} {2581} (\bibinfo {year} {2003})}\BibitemShut {NoStop}%
\bibitem [{\citenamefont {Klotsa}\ \emph {et~al.}(2005)\citenamefont {Klotsa},
  \citenamefont {R\"{o}mer},\ and\ \citenamefont {Turner}}]{Klotsa}%
  \BibitemOpen
  \bibfield  {author} {\bibinfo {author} {\bibfnamefont {D.}~\bibnamefont
  {Klotsa}}, \bibinfo {author} {\bibfnamefont {R.~A.}\ \bibnamefont
  {R\"{o}mer}}, \ and\ \bibinfo {author} {\bibfnamefont {M.~S.}\ \bibnamefont
  {Turner}},\ }\href@noop {} {\bibfield  {journal} {\bibinfo  {journal}
  {Biophys. Journal}\ }\textbf {\bibinfo {volume} {89}},\ \bibinfo {pages}
  {2187} (\bibinfo {year} {2005})}\BibitemShut {NoStop}%
\bibitem [{\citenamefont {Qu}\ \emph {et~al.}(2008)\citenamefont {Qu} \emph
  {et~al.}}]{Qu}%
  \BibitemOpen
  \bibfield  {author} {\bibinfo {author} {\bibfnamefont {Z.}~\bibnamefont {Qu}}
  \emph {et~al.},\ }\href@noop {} {\bibfield  {journal} {\bibinfo  {journal}
  {Frontiers of Physics in China}\ }\textbf {\bibinfo {volume} {3}},\ \bibinfo
  {pages} {349} (\bibinfo {year} {2008})}\BibitemShut {NoStop}%
\bibitem [{\citenamefont {Hjort}\ and\ \citenamefont
  {Stafstr\"{o}m}(2000)}]{Hjort}%
  \BibitemOpen
  \bibfield  {author} {\bibinfo {author} {\bibfnamefont {M.}~\bibnamefont
  {Hjort}}\ and\ \bibinfo {author} {\bibfnamefont {S.}~\bibnamefont
  {Stafstr\"{o}m}},\ }\href@noop {} {\bibfield  {journal} {\bibinfo  {journal}
  {Phys. Rev. B}\ }\textbf {\bibinfo {volume} {62}},\ \bibinfo {pages} {5245}
  (\bibinfo {year} {2000})}\BibitemShut {NoStop}%
\bibitem [{\citenamefont {Prigodin}\ and\ \citenamefont
  {Epstein}(2001)}]{PrigodinI}%
  \BibitemOpen
  \bibfield  {author} {\bibinfo {author} {\bibfnamefont {V.}~\bibnamefont
  {Prigodin}}\ and\ \bibinfo {author} {\bibfnamefont {A.}~\bibnamefont
  {Epstein}},\ }\href@noop {} {\bibfield  {journal} {\bibinfo  {journal}
  {Synth. Met.}\ }\textbf {\bibinfo {volume} {125}},\ \bibinfo {pages} {43}
  (\bibinfo {year} {2001})}\BibitemShut {NoStop}%
\bibitem [{\citenamefont {Prigodin}\ and\ \citenamefont
  {Roth}(1993)}]{PrigodinII}%
  \BibitemOpen
  \bibfield  {author} {\bibinfo {author} {\bibfnamefont {V.}~\bibnamefont
  {Prigodin}}\ and\ \bibinfo {author} {\bibfnamefont {S.}~\bibnamefont
  {Roth}},\ }\href@noop {} {\bibfield  {journal} {\bibinfo  {journal} {Synth.
  Metals}\ }\textbf {\bibinfo {volume} {53}},\ \bibinfo {pages} {237} (\bibinfo
  {year} {1993})}\BibitemShut {NoStop}%
\bibitem [{\citenamefont {Wang}\ and\ \citenamefont {Genack}(2011)}]{Wang}%
  \BibitemOpen
  \bibfield  {author} {\bibinfo {author} {\bibfnamefont {J.}~\bibnamefont
  {Wang}}\ and\ \bibinfo {author} {\bibfnamefont {Z.~Z.}\ \bibnamefont
  {Genack}},\ }\href@noop {} {\bibfield  {journal} {\bibinfo  {journal}
  {Nature}\ }\textbf {\bibinfo {volume} {471}},\ \bibinfo {pages} {345}
  (\bibinfo {year} {2011})}\BibitemShut {NoStop}%
\bibitem [{\citenamefont {Pascual}\ \emph {et~al.}(1995)\citenamefont {Pascual}
  \emph {et~al.}}]{Pascual}%
  \BibitemOpen
  \bibfield  {author} {\bibinfo {author} {\bibfnamefont {J.~I.}\ \bibnamefont
  {Pascual}} \emph {et~al.},\ }\href@noop {} {\bibfield  {journal} {\bibinfo
  {journal} {Science}\ }\textbf {\bibinfo {volume} {267}},\ \bibinfo {pages}
  {1793} (\bibinfo {year} {1995})}\BibitemShut {NoStop}%
\bibitem [{\citenamefont {Abrahams}\ \emph {et~al.}(1979)\citenamefont
  {Abrahams} \emph {et~al.}}]{Abrahams}%
  \BibitemOpen
  \bibfield  {author} {\bibinfo {author} {\bibfnamefont {E.}~\bibnamefont
  {Abrahams}} \emph {et~al.},\ }\href@noop {} {\bibfield  {journal} {\bibinfo
  {journal} {Phys. Rev. Lett.}\ }\textbf {\bibinfo {volume} {42}},\ \bibinfo
  {pages} {673} (\bibinfo {year} {1979})}\BibitemShut {NoStop}%
\bibitem [{\citenamefont {Abrahams}(2010)}]{50years}%
  \BibitemOpen
  \bibfield  {author} {\bibinfo {author} {\bibfnamefont {E.}~\bibnamefont
  {Abrahams}},\ }\href@noop {} {\emph {\bibinfo {title} {50 Years of Anderson
  Localization}}}\ (\bibinfo  {publisher} {World Scientific Publishing
  Company},\ \bibinfo {year} {2010})\BibitemShut {NoStop}%
\bibitem [{\citenamefont {Johnston}\ and\ \citenamefont
  {Kramer}(1986)}]{KramerB:1986}%
  \BibitemOpen
  \bibfield  {author} {\bibinfo {author} {\bibfnamefont {R.}~\bibnamefont
  {Johnston}}\ and\ \bibinfo {author} {\bibfnamefont {B.}~\bibnamefont
  {Kramer}},\ }\href@noop {} {\bibfield  {journal} {\bibinfo  {journal} {Z.
  Phys. B}\ }\textbf {\bibinfo {volume} {63}},\ \bibinfo {pages} {273}
  (\bibinfo {year} {1986})}\BibitemShut {NoStop}%
\bibitem [{\citenamefont {Jitomirskaya}(1999)}]{Jitomirskaya}%
  \BibitemOpen
  \bibfield  {author} {\bibinfo {author} {\bibfnamefont {S.~Y.}\ \bibnamefont
  {Jitomirskaya}},\ }\href@noop {} {\bibfield  {journal} {\bibinfo  {journal}
  {Ann. Math.}\ }\textbf {\bibinfo {volume} {150}},\ \bibinfo {pages} {1159}
  (\bibinfo {year} {1999})}\BibitemShut {NoStop}%
\bibitem [{\citenamefont {Avila}\ and\ \citenamefont {Damanik}(2008)}]{Avila}%
  \BibitemOpen
  \bibfield  {author} {\bibinfo {author} {\bibfnamefont {A.}~\bibnamefont
  {Avila}}\ and\ \bibinfo {author} {\bibfnamefont {D.}~\bibnamefont
  {Damanik}},\ }\href@noop {} {\bibfield  {journal} {\bibinfo  {journal}
  {Invent. Math.}\ }\textbf {\bibinfo {volume} {172}},\ \bibinfo {pages} {439}
  (\bibinfo {year} {2008})}\BibitemShut {NoStop}%
\bibitem [{\citenamefont {Flores}(1989)}]{Flores}%
  \BibitemOpen
  \bibfield  {author} {\bibinfo {author} {\bibfnamefont {J.~C.}\ \bibnamefont
  {Flores}},\ }\href@noop {} {\bibfield  {journal} {\bibinfo  {journal} {J.
  Phys.: Condens. Matter}\ }\textbf {\bibinfo {volume} {1}},\ \bibinfo {pages}
  {8471} (\bibinfo {year} {1989})}\BibitemShut {NoStop}%
\bibitem [{\citenamefont {Dunlap}\ \emph {et~al.}(1990)\citenamefont {Dunlap},
  \citenamefont {Wu},\ and\ \citenamefont {Phillips}}]{Phillips}%
  \BibitemOpen
  \bibfield  {author} {\bibinfo {author} {\bibfnamefont {D.~H.}\ \bibnamefont
  {Dunlap}}, \bibinfo {author} {\bibfnamefont {H.-L.}\ \bibnamefont {Wu}}, \
  and\ \bibinfo {author} {\bibfnamefont {P.~W.}\ \bibnamefont {Phillips}},\
  }\href@noop {} {\bibfield  {journal} {\bibinfo  {journal} {Phys. Rev. Lett.}\
  }\textbf {\bibinfo {volume} {65}},\ \bibinfo {pages} {88} (\bibinfo {year}
  {1990})}\BibitemShut {NoStop}%
\bibitem [{\citenamefont {de~Moura}\ and\ \citenamefont
  {Lyra}(1998)}]{DeMoura}%
  \BibitemOpen
  \bibfield  {author} {\bibinfo {author} {\bibfnamefont {F.~A. B.~F.}\
  \bibnamefont {de~Moura}}\ and\ \bibinfo {author} {\bibfnamefont {M.~L.}\
  \bibnamefont {Lyra}},\ }\href@noop {} {\bibfield  {journal} {\bibinfo
  {journal} {Phys. Rev. Lett.}\ }\textbf {\bibinfo {volume} {81}},\ \bibinfo
  {pages} {3735} (\bibinfo {year} {1998})}\BibitemShut {NoStop}%
\bibitem [{\citenamefont {Izrailev}\ \emph {et~al.}(2012)\citenamefont
  {Izrailev}, \citenamefont {Krokhin},\ and\ \citenamefont
  {Makarov}}]{Izrailev}%
  \BibitemOpen
  \bibfield  {author} {\bibinfo {author} {\bibfnamefont {F.~M.}\ \bibnamefont
  {Izrailev}}, \bibinfo {author} {\bibfnamefont {A.~A.}\ \bibnamefont
  {Krokhin}}, \ and\ \bibinfo {author} {\bibfnamefont {N.~M.}\ \bibnamefont
  {Makarov}},\ }\href@noop {} {\bibfield  {journal} {\bibinfo  {journal} {Phys.
  Reports}\ }\textbf {\bibinfo {volume} {512}},\ \bibinfo {pages} {125}
  (\bibinfo {year} {2012})}\BibitemShut {NoStop}%
\bibitem [{\citenamefont {Billy}\ \emph {et~al.}(2008)\citenamefont {Billy}
  \emph {et~al.}}]{Billy}%
  \BibitemOpen
  \bibfield  {author} {\bibinfo {author} {\bibfnamefont {J.}~\bibnamefont
  {Billy}} \emph {et~al.},\ }\href@noop {} {\bibfield  {journal} {\bibinfo
  {journal} {Nature}\ }\textbf {\bibinfo {volume} {543}},\ \bibinfo {pages}
  {891} (\bibinfo {year} {2008})}\BibitemShut {NoStop}%
\bibitem [{\citenamefont {Roati}\ \emph {et~al.}(2008)\citenamefont {Roati}
  \emph {et~al.}}]{Inguscio}%
  \BibitemOpen
  \bibfield  {author} {\bibinfo {author} {\bibfnamefont {G.}~\bibnamefont
  {Roati}} \emph {et~al.},\ }\href@noop {} {\bibfield  {journal} {\bibinfo
  {journal} {Nature}\ }\textbf {\bibinfo {volume} {453}},\ \bibinfo {pages}
  {895} (\bibinfo {year} {2008})}\BibitemShut {NoStop}%
\bibitem [{\citenamefont {Clark}\ \emph {et~al.}(2010)\citenamefont {Clark}
  \emph {et~al.}}]{Saw1}%
  \BibitemOpen
  \bibfield  {author} {\bibinfo {author} {\bibfnamefont {K.}~\bibnamefont
  {Clark}} \emph {et~al.},\ }\href@noop {} {\bibfield  {journal} {\bibinfo
  {journal} {Nature Nanotech.}\ }\textbf {\bibinfo {volume} {5}},\ \bibinfo
  {pages} {261} (\bibinfo {year} {2010})}\BibitemShut {NoStop}%
\bibitem [{\citenamefont {Dilullo}\ \emph {et~al.}(2012)\citenamefont {Dilullo}
  \emph {et~al.}}]{Saw2}%
  \BibitemOpen
  \bibfield  {author} {\bibinfo {author} {\bibfnamefont {A.}~\bibnamefont
  {Dilullo}} \emph {et~al.},\ }\href@noop {} {\bibfield  {journal} {\bibinfo
  {journal} {Nano. Lett.}\ }\textbf {\bibinfo {volume} {12}},\ \bibinfo {pages}
  {3174} (\bibinfo {year} {2012})}\BibitemShut {NoStop}%
\bibitem [{\citenamefont {Novoselov}\ and\ \citenamefont
  {Geim}(2005)}]{Novoselov}%
  \BibitemOpen
  \bibfield  {author} {\bibinfo {author} {\bibfnamefont {K.~S.}\ \bibnamefont
  {Novoselov}}\ and\ \bibinfo {author} {\bibfnamefont {A.~K.}\ \bibnamefont
  {Geim}},\ }\href@noop {} {\bibfield  {journal} {\bibinfo  {journal} {PNAS}\
  }\textbf {\bibinfo {volume} {102}},\ \bibinfo {pages} {10451} (\bibinfo
  {year} {2005})}\BibitemShut {NoStop}%
\bibitem [{\citenamefont {Lau}\ \emph {et~al.}(2012)\citenamefont {Lau},
  \citenamefont {Bao},\ and\ \citenamefont {Jr.}}]{Lau}%
  \BibitemOpen
  \bibfield  {author} {\bibinfo {author} {\bibfnamefont {C.~N.}\ \bibnamefont
  {Lau}}, \bibinfo {author} {\bibfnamefont {W.}~\bibnamefont {Bao}}, \ and\
  \bibinfo {author} {\bibfnamefont {J.~V.}\ \bibnamefont {Jr.}},\ }\href@noop
  {} {\bibfield  {journal} {\bibinfo  {journal} {Materials Today}\ }\textbf
  {\bibinfo {volume} {15}},\ \bibinfo {pages} {238} (\bibinfo {year}
  {2012})}\BibitemShut {NoStop}%
\bibitem [{\citenamefont {Wang}\ \emph {et~al.}(2011)\citenamefont {Wang} \emph
  {et~al.}}]{Wang-Y}%
  \BibitemOpen
  \bibfield  {author} {\bibinfo {author} {\bibfnamefont {Y.}~\bibnamefont
  {Wang}} \emph {et~al.},\ }\href@noop {} {\bibfield  {journal} {\bibinfo
  {journal} {ACS Nano}\ }\textbf {\bibinfo {volume} {5}},\ \bibinfo {pages}
  {3645} (\bibinfo {year} {2011})}\BibitemShut {NoStop}%
\bibitem [{\citenamefont {Russ}\ \emph {et~al.}(1999)\citenamefont {Russ} \emph
  {et~al.}}]{Russ}%
  \BibitemOpen
  \bibfield  {author} {\bibinfo {author} {\bibfnamefont {S.}~\bibnamefont
  {Russ}} \emph {et~al.},\ }\href@noop {} {\bibfield  {journal} {\bibinfo
  {journal} {Physica A}\ }\textbf {\bibinfo {volume} {266}},\ \bibinfo {pages}
  {492} (\bibinfo {year} {1999})}\BibitemShut {NoStop}%
\bibitem [{\citenamefont {Croy}\ \emph {et~al.}(2011)\citenamefont {Croy},
  \citenamefont {Cain},\ and\ \citenamefont {Schreiber}}]{CroyA}%
  \BibitemOpen
  \bibfield  {author} {\bibinfo {author} {\bibfnamefont {A.}~\bibnamefont
  {Croy}}, \bibinfo {author} {\bibfnamefont {P.}~\bibnamefont {Cain}}, \ and\
  \bibinfo {author} {\bibfnamefont {M.}~\bibnamefont {Schreiber}},\ }\href@noop
  {} {\bibfield  {journal} {\bibinfo  {journal} {Eur. Phys. J. B}\ }\textbf
  {\bibinfo {volume} {82}},\ \bibinfo {pages} {2012} (\bibinfo {year}
  {2011})}\BibitemShut {NoStop}%
\bibitem [{\citenamefont {Izrailev}\ and\ \citenamefont
  {Krokhin}(1999)}]{Krokhin}%
  \BibitemOpen
  \bibfield  {author} {\bibinfo {author} {\bibfnamefont {F.~M.}\ \bibnamefont
  {Izrailev}}\ and\ \bibinfo {author} {\bibfnamefont {A.~A.}\ \bibnamefont
  {Krokhin}},\ }\href@noop {} {\bibfield  {journal} {\bibinfo  {journal} {Phys.
  Rev. Lett.}\ }\textbf {\bibinfo {volume} {82}},\ \bibinfo {pages} {4062}
  (\bibinfo {year} {1999})}\BibitemShut {NoStop}%
\bibitem [{\citenamefont {Schomerus}\ and\ \citenamefont
  {Titov}(2003)}]{Schomerus}%
  \BibitemOpen
  \bibfield  {author} {\bibinfo {author} {\bibfnamefont {H.}~\bibnamefont
  {Schomerus}}\ and\ \bibinfo {author} {\bibfnamefont {M.}~\bibnamefont
  {Titov}},\ }\href@noop {} {\bibfield  {journal} {\bibinfo  {journal} {Phys.
  Rev. B}\ }\textbf {\bibinfo {volume} {67}},\ \bibinfo {pages} {100201}
  (\bibinfo {year} {2003})}\BibitemShut {NoStop}%
\bibitem [{\citenamefont {Deych}\ \emph {et~al.}(2003)\citenamefont {Deych}
  \emph {et~al.}}]{Deych}%
  \BibitemOpen
  \bibfield  {author} {\bibinfo {author} {\bibfnamefont {L.}~\bibnamefont
  {Deych}} \emph {et~al.},\ }\href@noop {} {\bibfield  {journal} {\bibinfo
  {journal} {Phys. Rev. Lett.}\ }\textbf {\bibinfo {volume} {91}},\ \bibinfo
  {pages} {96601} (\bibinfo {year} {2003})}\BibitemShut {NoStop}%
\bibitem [{\citenamefont {Kravtsov}\ and\ \citenamefont
  {Yudson}(2012)}]{Kravtsov}%
  \BibitemOpen
  \bibfield  {author} {\bibinfo {author} {\bibfnamefont {V.~E.}\ \bibnamefont
  {Kravtsov}}\ and\ \bibinfo {author} {\bibfnamefont {V.~I.}\ \bibnamefont
  {Yudson}},\ }\href@noop {} {\bibfield  {journal} {\bibinfo  {journal} {arXiv
  cond-mat}\ ,\ \bibinfo {pages} {1208.4789}} (\bibinfo {year}
  {2012})}\BibitemShut {NoStop}%
\bibitem [{\citenamefont {Weinrib}\ and\ \citenamefont
  {Halperin}(1983)}]{Weinrib}%
  \BibitemOpen
  \bibfield  {author} {\bibinfo {author} {\bibfnamefont {A.}~\bibnamefont
  {Weinrib}}\ and\ \bibinfo {author} {\bibfnamefont {B.}~\bibnamefont
  {Halperin}},\ }\href@noop {} {\bibfield  {journal} {\bibinfo  {journal}
  {Phys. Rev. B}\ }\textbf {\bibinfo {volume} {27}},\ \bibinfo {pages} {413}
  (\bibinfo {year} {1983})}\BibitemShut {NoStop}%
\bibitem [{\citenamefont {Sandler}\ \emph {et~al.}(2003)\citenamefont
  {Sandler}, \citenamefont {Maei},\ and\ \citenamefont {Kondev}}]{Sandler}%
  \BibitemOpen
  \bibfield  {author} {\bibinfo {author} {\bibfnamefont {N.}~\bibnamefont
  {Sandler}}, \bibinfo {author} {\bibfnamefont {H.~R.}\ \bibnamefont {Maei}}, \
  and\ \bibinfo {author} {\bibfnamefont {J.}~\bibnamefont {Kondev}},\
  }\href@noop {} {\bibfield  {journal} {\bibinfo  {journal} {Phys. Rev. B}\
  }\textbf {\bibinfo {volume} {68}},\ \bibinfo {pages} {205315} (\bibinfo
  {year} {2003})}\BibitemShut {NoStop}%
\bibitem [{\citenamefont {Petersen}\ and\ \citenamefont {Sandler}()}]{Greg2}%
  \BibitemOpen
  \bibfield  {author} {\bibinfo {author} {\bibfnamefont {G.}~\bibnamefont
  {Petersen}}\ and\ \bibinfo {author} {\bibfnamefont {N.}~\bibnamefont
  {Sandler}},\ }\href@noop {} {\bibinfo  {journal} {unpublished}\ }\BibitemShut
  {NoStop}%
\bibitem [{\citenamefont {de~Moura}()}]{Greg3}%
  \BibitemOpen
\bibfield  {journal} {  }\bibfield  {author} {\bibinfo {author} {\bibfnamefont
  {F.}~\bibnamefont {de~Moura}},\ }\href@noop {} {\bibinfo  {journal} {private
  communication}\ }\BibitemShut {NoStop}%
\bibitem [{\citenamefont {Shapiro}(1987)}]{Shapiro}%
  \BibitemOpen
\bibfield  {journal} {  }\bibfield  {author} {\bibinfo {author} {\bibfnamefont
  {B.}~\bibnamefont {Shapiro}},\ }\href@noop {} {\bibfield  {journal} {\bibinfo
   {journal} {Philos. Mag.}\ }\textbf {\bibinfo {volume} {56}},\ \bibinfo
  {pages} {1031} (\bibinfo {year} {1987})}\BibitemShut {NoStop}%
\bibitem [{\citenamefont {Beenakker}(1997)}]{Beenakker}%
  \BibitemOpen
  \bibfield  {author} {\bibinfo {author} {\bibfnamefont {C.~W.~J.}\
  \bibnamefont {Beenakker}},\ }\href@noop {} {\bibfield  {journal} {\bibinfo
  {journal} {Rev. Mod. Phys.}\ }\textbf {\bibinfo {volume} {69}},\ \bibinfo
  {pages} {731} (\bibinfo {year} {1997})}\BibitemShut {NoStop}%
\bibitem [{\citenamefont {Lake}\ \emph {et~al.}(1997)\citenamefont {Lake} \emph
  {et~al.}}]{LakeR}%
  \BibitemOpen
  \bibfield  {author} {\bibinfo {author} {\bibfnamefont {R.}~\bibnamefont
  {Lake}} \emph {et~al.},\ }\href@noop {} {\bibfield  {journal} {\bibinfo
  {journal} {J. Appl. Phys.}\ }\textbf {\bibinfo {volume} {81}},\ \bibinfo
  {pages} {7845} (\bibinfo {year} {1997})}\BibitemShut {NoStop}%
\bibitem [{\citenamefont {Arrachea}(2007)}]{Arrachea}%
  \BibitemOpen
  \bibfield  {author} {\bibinfo {author} {\bibfnamefont {L.}~\bibnamefont
  {Arrachea}},\ }\href@noop {} {\bibfield  {journal} {\bibinfo  {journal}
  {Phys. Rev. B}\ }\textbf {\bibinfo {volume} {75}},\ \bibinfo {pages} {035319}
  (\bibinfo {year} {2007})}\BibitemShut {NoStop}%
\bibitem [{\citenamefont {Klimeck}(2010)}]{RGF}%
  \BibitemOpen
  \bibfield  {author} {\bibinfo {author} {\bibfnamefont {G.}~\bibnamefont
  {Klimeck}},\ }\href@noop {} {\bibfield  {journal} {\bibinfo  {journal}
  {Lecture: Recursive Green Function Algorithm}\ ,\ \bibinfo {pages}
  {http://nanohub.org/resources/8388}} (\bibinfo {year} {2010})}\BibitemShut
  {NoStop}%
\bibitem [{\citenamefont {Derrida}\ and\ \citenamefont
  {Gardner}(1984)}]{Derrida}%
  \BibitemOpen
  \bibfield  {author} {\bibinfo {author} {\bibfnamefont {B.}~\bibnamefont
  {Derrida}}\ and\ \bibinfo {author} {\bibfnamefont {E.}~\bibnamefont
  {Gardner}},\ }\href@noop {} {\bibfield  {journal} {\bibinfo  {journal} {J.
  Physique}\ }\textbf {\bibinfo {volume} {45}},\ \bibinfo {pages} {1283}
  (\bibinfo {year} {1984})}\BibitemShut {NoStop}%
\bibitem [{\citenamefont {Thouless}(1974)}]{Thouless}%
  \BibitemOpen
  \bibfield  {author} {\bibinfo {author} {\bibfnamefont {J.~J.}\ \bibnamefont
  {Thouless}},\ }\href@noop {} {\bibfield  {journal} {\bibinfo  {journal}
  {Physics Reports}\ }\textbf {\bibinfo {volume} {13}},\ \bibinfo {pages} {93}
  (\bibinfo {year} {1974})}\BibitemShut {NoStop}%
\bibitem [{\citenamefont {Roman}\ and\ \citenamefont {Wiecko}(1986)}]{Roman1}%
  \BibitemOpen
  \bibfield  {author} {\bibinfo {author} {\bibfnamefont {E.}~\bibnamefont
  {Roman}}\ and\ \bibinfo {author} {\bibfnamefont {C.}~\bibnamefont {Wiecko}},\
  }\href@noop {} {\bibfield  {journal} {\bibinfo  {journal} {Z. Phys. B}\
  }\textbf {\bibinfo {volume} {62}},\ \bibinfo {pages} {163} (\bibinfo {year}
  {1986})}\BibitemShut {NoStop}%
\bibitem [{\citenamefont {Roman}(1986)}]{Roman2}%
  \BibitemOpen
  \bibfield  {author} {\bibinfo {author} {\bibfnamefont {E.}~\bibnamefont
  {Roman}},\ }\href@noop {} {\bibfield  {journal} {\bibinfo  {journal} {J.
  Phys. C: Solid State Phys.}\ }\textbf {\bibinfo {volume} {19}},\ \bibinfo
  {pages} {285} (\bibinfo {year} {1986})}\BibitemShut {NoStop}%
\bibitem [{\citenamefont {Evers}\ and\ \citenamefont {Mirlin}(2008)}]{Evers}%
  \BibitemOpen
  \bibfield  {author} {\bibinfo {author} {\bibfnamefont {F.}~\bibnamefont
  {Evers}}\ and\ \bibinfo {author} {\bibfnamefont {A.}~\bibnamefont {Mirlin}},\
  }\href@noop {} {\bibfield  {journal} {\bibinfo  {journal} {Rev. Mod. Phy.}\
  }\textbf {\bibinfo {volume} {80}},\ \bibinfo {pages} {1355} (\bibinfo {year}
  {2008})}\BibitemShut {NoStop}%
\end{thebibliography}%

\end{document}